\documentclass[12pt,epsf,showkeys]{article}

\usepackage{graphicx,float}
\usepackage{mathrsfs,array,multirow}
\usepackage{amstext}
\usepackage{subfigure}
\usepackage{bm,latexsym,amsmath,amsfonts,amssymb}
\usepackage[usenames,dvipsnames]{color}
\usepackage{color}
\usepackage[colorlinks=true,linkcolor=blue]{hyperref}
\usepackage{soul}
\usepackage{epsfig}
\usepackage{subfigure}	

\newcommand{\be}{\begin{equation}}

\newcommand{\ee}{\end{equation}}

\newcommand{\ba}{\begin{array}}

\newcommand{\ea}{\end{array}}

\begin{document}
\begin{titlepage}
\vspace{.5in}
\begin{flushright}
%CQUeST-2021-0662
\end{flushright}
\vspace{0.5cm}

\begin{center}
{\Large\bf Gravitational waves from the vacuum decay with LISA}\\
\vspace{.4in}

  {$\mbox{Bum-Hoon Lee}^{\S \dag\copyright}$}\footnote{\it email: bhl@sogang.ac.kr},\,\,
  {$\mbox{Wonwoo Lee}^{\S}$}\footnote{\it email: warrior@sogang.ac.kr},\, \,
  {$\mbox{Dong-han Yeom}^{\P}$}\footnote{\it email: innocent.yeom@gmail.com},\, \,
  {$\mbox{Lu Yin}^{\S\dag}$}\footnote{\it email: yinlu@sogang.ac.kr}\\

\vspace{.3in}

{\small \S \it Center for Quantum Spacetime, Sogang University, Seoul 04107, Korea}\\
{\small \dag \it Department of Physics, Sogang University, Seoul 04107, Korea}\\
{\small \copyright \it Asia Pacific Center for Theoretical Physics, Postech, Pohang 37673, Korea}\\
{\small \pounds \it Department of Physics Education, Pusan National University, Busan 46241, Korea}\\
{\small \ddag \it Research Center for Dielectric and Advanced Matter Physics, Pusan National University, Busan 46241, Korea}\\

\vspace{.5in}
\end{center}
\begin{center}
{\large\bf Abstract}
\end{center}
\begin{center}
\begin{minipage}{4.75in}

{\small \,\,\,\,
We investigate the gravitational wave spectrum originating from the cosmological
first-order phase transition. We compare two models: one is a scalar field model without gravitation,
while the other is a scalar field model with gravitation.
Based on the sensitivity curves of the LISA space-based interferometer on the stochastic
gravitational-wave background, we compare the difference between the gravitational wave spectra of
the former and the latter cases obtained from the bubble collision process.
In particular,  we numerically calculated the speed of the bubble wall before collision for the two models.
 We demonstrate that the difference between
the amplitudes of these spectra can clearly distinguish between the two models.
We expect that the LISA with Signal to Noise Ratio =10 could observe the spectrum as the fast first-order phase transition.}
\end{minipage}
\end{center}
\end{titlepage}

\newpage

\section{ Introduction \label{sec1}}

It was reported that gravitational waves (GWs) from a binary system comprising two black holes were directly detected,
in which the dominant portion of the GWs was emitted just before the collision \cite{Abbott:2016nmj, Abbott:2017vtc}.
We have now opened a new window toward an unknown area that has not yet been investigated.
Furthermore, we have a new opportunity to observe the past event of the universe, which took place
in the regime with the strong gravitational field. 
Recently, GW detection has become
a new challenge \cite{AmaroSeoane:2012km, Caprini:2019egz}, especially both in astrophysics and cosmology.
Some novel topics are garnering more attention, including the GWs obtained from the pre-inflation era,
 inflation era \cite{Starobinsky:1979ty, Koh:2004ez, Bartolo:2016ami, Gong:2017qlj, Koh:2018qcy,Huang:2019lgd},
 gravitational waves that could originate from the electroweak phase transition era
\cite{Binetruy:2012ze, Caprini:2015zlo, Cai:2017tmh, Caprini:2019egz, Jinno:2017fby,Schmitz:2020syl},
and others \cite{Vilenkin:1981bx, Barenboim:2016mjm, Lewicki:2019gmv, Guo:2020grp, Wang:2020jrd}.

The origins of the stochastic gravitational waves background (SGWB) have different types of possibility. The first one is the SGWB induced nonlinearly by curvature perturbations, which is strongly related to primordial non-Gaussianity\cite{Wang:2019kaf}. The second possibility is that of non-perturbative effects, such as sound speed resonance \cite{Cai:2019jah, Cai:2020ovp} or heave field resonance \cite{Cai:2021yvq, Zhou:2020kkf}, which could significantly enhance cosmological gravitational waves at a certain frequency band.
The cosmological first-order phase transition was first studied by the scalar field model
in \cite{Kobzarev:1974cp, Coleman:1977py}. In addition, the effects of gravity were considered
in \cite{Coleman:1980aw, Parke:1982pm, Lee:1987qc, Lee:2008hz, Lee:2009bp, Zhang:2013pna,Joti:2017fwe, Wong:2017wsw, Branchina:2018xdh, Markkanen:2018pdo,Bramberger:2019mkv}. The authors studied the nucleation
process of a vacuum bubble with $O(4)$ symmetry in the Euclidean space.  Owing to the tunneling,
the inside is in a lower vacuum energy state (or the broken phase),
the outside is in a higher vacuum energy state (or the symmetric phase),
and the transition region becomes a bubble wall. After the materialization
of the vacuum bubble, the analytic continuation is conducted from Lorentzian to Euclidean signatures;
eventually, the bubble expands over spacetime. Immediately after that, the procedure was applied to
inflationary scenarios \cite{Guth:1980zm, Sato:1980yn, Hawking:1981fz, La:1989za}.
 In addition, the GW can be affected if multiple vacua were involved \cite{Imtiaz:2018dfn, Zhou:2020stj}.

% (negative sentiment) , although now lots of the first-order phase transition models of the primordial inflation have been ruled out.

In previous investigations, the authors obtained the analytic forms for the nucleation rate
of a true vacuum bubble and its radius. When gravity is considered, the nucleation rate
increases and the radius decreases, compared to those without gravity when the transition occurs
from the de Sitter space to the flat Minkowski space \cite{Coleman:1980aw}. As illustrated in \cite{Lee:2008hz},
the nucleation rate could increase up to four times, provided it could have the physical meaning.

The phase transition due to the thermal effect was studied in \cite{Linde:1980tt, Linde:1981zj}.
In this study, the Euclidean time becomes the inverse of the temperature; except for the time in which
an $O(3)$-symmetric vacuum bubble is generated, and it evolves in the Lorentzian spacetime.
It appears difficult to simultaneously obtain the analytic forms that include both the thermal and gravity
effects. Some studies have been conducted on $O(3)$-symmetric bubble with gravitation
\cite{Berezin:1987ea, Berezin:1990qs, Gregory:2020cvy}. We note that the nucleation rate of
$O(4)$-symmetric vacuum bubbles is more dominant than those with $O(3)$ symmetry \cite{Coleman:1977th}.

In the thermal history of the Universe with the Standard Model of particle physics, the electroweak
phase transition was not the first-order but the smooth crossover. However, we expect that
the baryon asymmetry in the Universe could explain why the first-order phase
transition must occur \cite{Linde:1978px, Kolb:1983ni, Kuzmin:1985mm, Grojean:2004xa, Bodeker:2004ws, Morrissey:2012db, Xie:2020bkl,Xie:2020wzn, Hindmarsh:2020hop}.
Typically, when the electroweak phase transition was in the first-order,
it was applied with its temperature, ignoring the influence of gravity.
Although we cannot obtain correct analytic forms
of the nucleation rate and the bubble radius with $O(4)$-symmetry at the time of the electroweak phase transition
with the finite temperature when considering the gravity, we can expect that the gravity effect will
trigger more possibilities of the GW detection via space-based laser interferometry, such as
 Laser Interferometer Space Antenna (LISA).

LISA is a space probe that has been proposed to detect and accurately measure the GWs in the sub-Hz region
from astronomical and cosmological sources.
The LISA mission is designed to directly observe the GWs via laser interferometry.
If an energy density more than $10^{-5}$ of the total energy density is converted to the
gravitational radiation at the time when that is generated, LISA is sufficiently sensitive
to detect the cosmological stochastic GW background that occurs in the range of energy from
$0.1$ TeV to $1000$ TeV~\cite{AmaroSeoane:2012km, Binetruy:2012ze}.
We focus on the detection of the GWs by the bubble collision that LISA could
potentially probe, which was generated by the first-order phase transition
(at the electroweak scale) in the early universe.

In this study, we introduce the model-dependent parameters of the GW spectrum and solely compare the parameters of the scalar field \cite{Lewicki:2019gmv} and those of the gravity cases.
We describe a bubble nucleation rate and its radius when the gravity effect is strong,
and demonstrate the collision phenomenon of two vacuum bubbles
by adopting double-null formalism in Section~\ref{sec2}. We compute the GW spectrum.
We compare the GW spectrum from the two models with the LISA sensitivity
in Section~\ref{sec3}. Finally, we summarize our results and discuss relevant matters in Section~\ref{sec4}.

\section{ Model-dependent parameters in gravitational wave spectrum \label{sec2}}

Generally, when a bubble wall eventually collides with another bubble wall, the kinetic energy disappears and
converts into GWs and thermal particles. The collision of vacuum bubbles was considered for
the first time in \cite{Hawking:1982ga} without gravity, in which the collision of the two bubbles was studied.
When gravity was considered, it was studied in \cite{Wu:1984eda}. In the bubble wall collision process,
the scalar field repeatedly oscillates backward and forward between false and true vacua on the potential,
while reducing the kinetic energy of the expanding bubble wall. Eventually, as the collision process gradually ends,
it will settle and disappear within a true vacuum; the first-order phase transition is then completed.
By adopting numerical computations, it was investigated in full
general relativity \cite{Johnson:2011wt, Hwang:2012pj, Hwang:2014cqa, Wainwright:2013lea}.

The GW spectrum from the bubble collision depends on four main parameters \cite{Caprini:2015zlo,Wang:2020zlf}:
the temperature $T_*$,   bubble wall speed $v_w$,  strength parameter $\alpha$, and  transition rate $\beta$.

 $T_*$ is the temperature when GWs are produced from bubble collisions,
and the nucleation temperature $T_n \approx T_*$ is adopted for the typical transition without significant reheating.
We follow this assumption in this study and take the value $T_* \approx 100$ GeV.

The $v_w$ parameter represents the bubble wall speed when bubble walls collide.
If the bubble expands at the mean-field level, the bubble wall can accelerate without any bound, which implies that
the maximum velocity ($v_w$) of the runaway bubble wall will quickly approach the speed of light.
We will consider the numerical value of $v_w$.
%in Table~\ref{tab:2} and Figure~\ref{fg:v}.

The strength parameter $\alpha$ is the ratio between the vacuum energy density $\rho_{\mathrm{vac}}$
and that of the radiation bath $\rho_{\mathrm{rad}}^*$, where $\rho_{\mathrm{rad}}^*= g_*\pi^2 T_*^4/30$.
Under the $T_*$ parameter, the $g_*$ represents the relativistic degrees of freedom in plasma.
We will assume that $\alpha \approx 1$ and $g_* \approx 106.75$ in the following discussion.

The transition rate of the vacuum bubble, i.e.,
\begin{eqnarray}
\beta = -\left.\frac{d B}{dt}\right|_{t=t_*}, \nonumber
\end{eqnarray}
will differ  between the scalar field only and that with gravity cases, where $B$ denotes the Euclidean action, which will be considered in the next Section.
{ In the phase transition by the thermal effect,
	the vacuum bubble nucleation rate is not a time-independent constant.
	Therefore, $\beta$ is non-zero, which is given by the Lorentzian time derivative
	of the log function of the decay rate.
	Because we do not know the analytic form of the nucleation rate with the $O(3)$ symmetry
	when considering both gravitation and temperature simultaneously,
	we adopt the difference between the cases with and without gravitation.
	We anticipate this difference to be determined by $\beta$ even with the temperature effect.}

The ratio of the energy density inside and outside the bubble also plays an important role in the
GW spectrum, which is given by $\kappa_{\phi} = \rho_{\phi}/\rho_{\mathrm{vac}}$.
% where {\color{red}$\rho_{\phi}$ and $\rho_{\mathrm{vac}}$ are the case of the scalar field only and
%the case of the scalar field with gravity, respectively. ?}
Because we do not calculate the interaction
of the scalar field with plasma in this work, the energy of the false vacuum can turn into the true vacuum without loss.
Hence, $\kappa_{\phi}$ will be considered as $1$.

\subsection{Decay of a metastable state and a vacuum bubble nucleation \label{sec2-1}}

We consider the action
\begin{eqnarray}
S= \int_{\mathcal M} \sqrt{-g} d^4 x \left[ \frac{R}{2\kappa}
-\frac{1}{2}\nabla_{\alpha}\phi \nabla^{\alpha}\phi
-V(\phi)\right] + S_{\mathrm{bd}} \,,
\label{f-action}
\end{eqnarray}
{where $\kappa \equiv 8\pi G$, $m_p =\frac{1}{\sqrt{\kappa}}\simeq 2.44 \times 10^{18} $GeV represents the
	reduced Planck mass, $g\equiv det g_{\mu\nu}$}, and the second term on the right-hand
side is the boundary term \cite{York:1972sj, Gibbons:1976ue}. The potential $V(\phi)$ has two non-degenerate
minima defined as  \cite{Hwang:2012pj}:
\begin{eqnarray}
\label{eq:3}
V(\phi)=\frac{6 V_{\mathrm{f}}}{5\phi_{\mathrm{f}}}\int_{0}^{\phi}\frac{\bar{\phi}}{\phi_{\mathrm{f}}}
\left(\frac{\bar{\phi}}{\phi_{\mathrm{f}}}-0.55\right)\left(\frac{\bar{\phi}}{\phi_{\mathrm{f}}}-1\right)d\bar{\phi} \,,
\end{eqnarray}
where $V_{\mathrm{f}}$ is the false vacuum energy density and $\phi_{\mathrm{f}}$ denotes the field value of that vacuum state.
The free parameter $S_{\mathrm{f}}(=\sqrt{\kappa/2}\phi_{\mathrm{f}})$ can be considered as the field distance between
the true and false vacuum states.
{ For the electroweak phase transition that occurred at $T \thicksim 100GeV$, } $S_{\mathrm{f}} $ has a value of about $ 10^{-16}$.
 It is easy to understand that tunneling will be difficult to occur if $S_{\mathrm{f}}$ is too excessively large.
{  In this study, we introduce  new dimensionless scaling variables as ${\bar \phi}\equiv \phi/{10^{-15}m_p}$
   and ${\bar V}\equiv V/(10^{-15}m_p)^4$, and then adopt the conventions $c=\hbar=1$ for simplicity.}

\begin{figure}
\begin{center}
{\includegraphics[width=3. in]{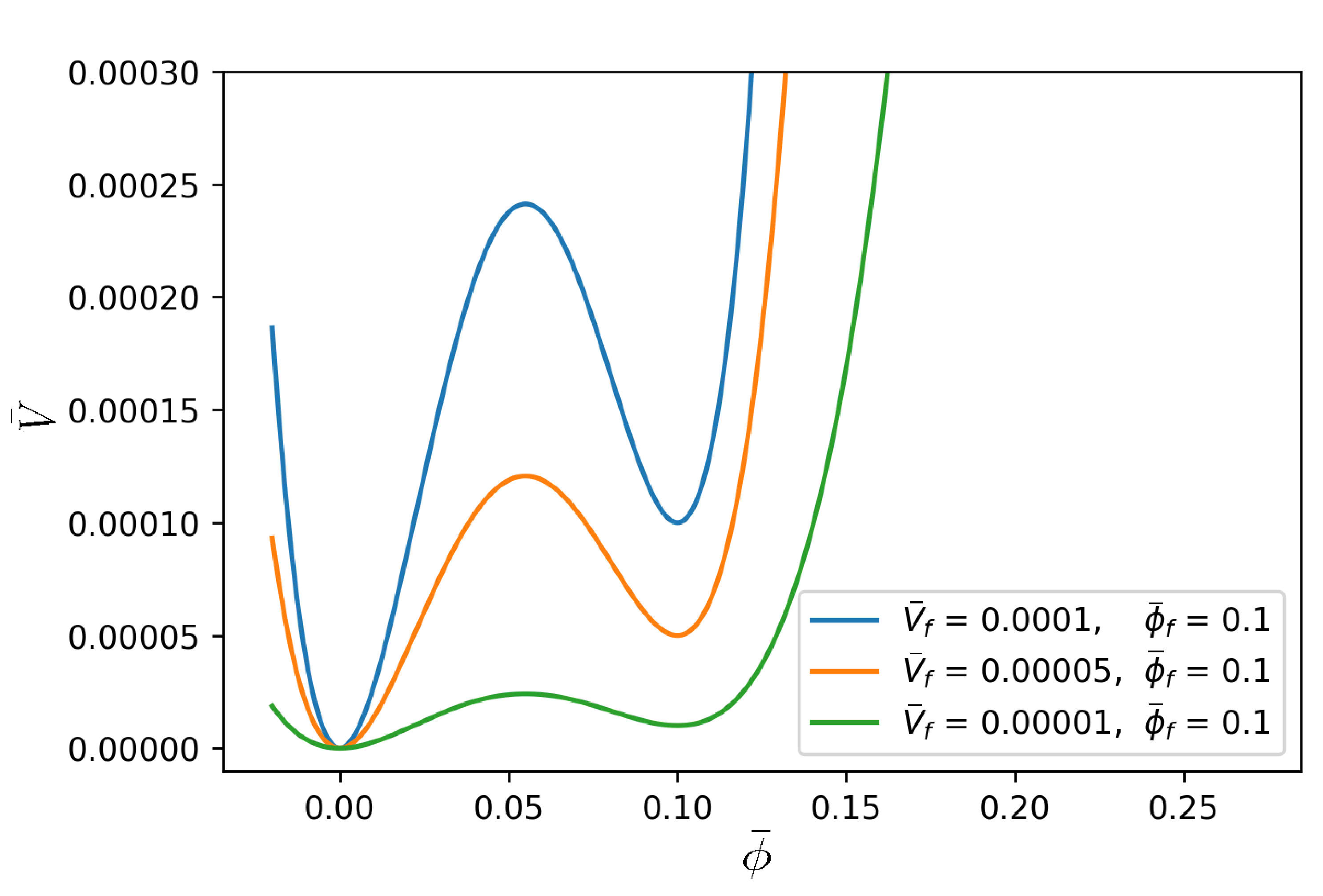}}
{\includegraphics[width=3. in]{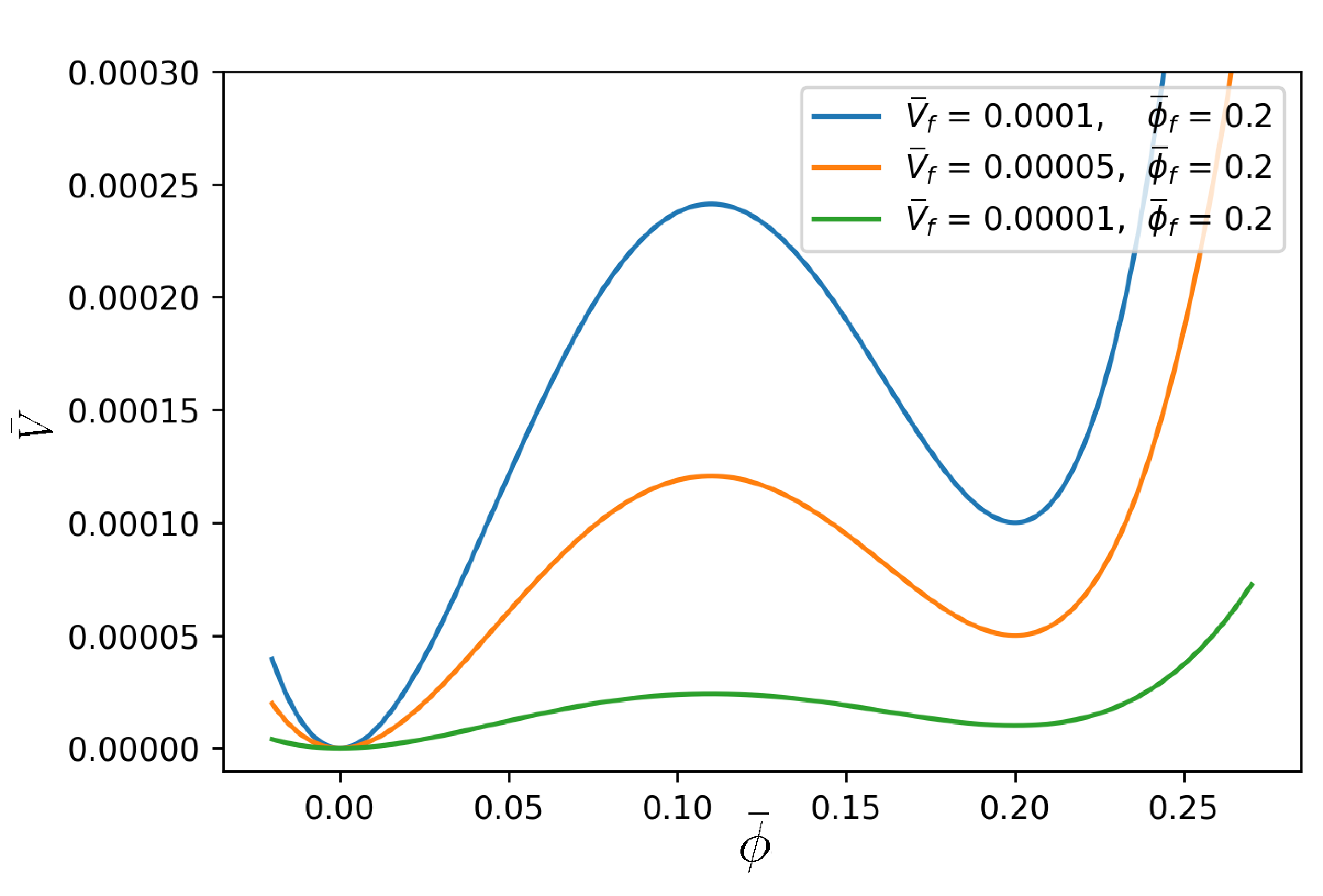}}
\end{center}
\caption{\footnotesize{(color online).
{ Potential with different value of ${\bar \phi}$ and ${\bar V}$ from Eq.~(\ref{eq:3}).}}}
\label{fg:3}
\end{figure}
Figure~\ref{fg:3} presents the potentials adopted in the numerical computations.
{The left figure presents ${\bar \phi}_{\mathrm{f}}=0.1 $, and ${\bar V}_{\mathrm{f}}$ corresponds to $0.0001$,
	$0.00005$ and $0.00001$,
	respectively. The right figure presents ${\bar \phi}_{\mathrm{f}}=0.2$ and ${\bar V}_{\mathrm{f}}$ exhibits the same values.}
%The left figure has $S_{\mathrm{f}}=0.1$, $V_{\mathrm{f}}$s correspond
%to $0.0001$, $0.00005$, and $0.00001$, respectively.
%The right figure has $S_{\mathrm{f}}=0.2$ and $V_{\mathrm{f}}$s with the same values.

{Now, we describe the vacuum bubble nucleation rate. }
The vacuum bubble nucleation rate per unit volume and unit time is semi-classically represented as
\begin{eqnarray}
\Gamma \simeq A \; e^{-B},
\end{eqnarray}
where the exponent $B$ is the difference between the Euclidean action of the instanton solution
$(S^{\mathrm{in}}_{\mathrm{E}})$ and the background action itself $(S^{\mathrm{b}}_{\mathrm{E}})$, i.e.,
\begin{eqnarray}
B = S^{\mathrm{in}}_{\mathrm{E}} - S^{\mathrm{b}}_{\mathrm{E}}.
\end{eqnarray}
The prefactor $A$ originates from the first-order quantum correction \cite{Callan:1977pt, Strumia:1999fv, Baacke:2003uw, Dunne:2005rt}.
One can further take the $O(4)$ symmetry for both the scalar field and the spacetime metric expecting
its dominant contribution \cite{Coleman:1977th}. The Euclidean metric is given by
\begin{eqnarray}
ds^2 = d\eta^2 + \rho^2(\eta)\left(d\chi^2 + \sin^2\chi d\Omega^2_2\right) \,.
\label{gemetric}
\end{eqnarray}

According to \cite{Coleman:1977py, Coleman:1980aw, Lee:2008hz}, the thin-wall approximation
scheme can be assumed to evaluate $B$. The validity of the approximation has been examined \cite{Samuel:1991mz, Samuel:1991dy, Brown:2017cca}.
In this approximation, the Euclidean action can be divided into three parts: $B = B_{\mathrm{in}} + B_{\mathrm{wall}} + B_{\mathrm{out}}$.
The configuration of the outside the wall will not be changed before and after the bubble formation.
Therefore, $B_{\mathrm{out}}=0$. Hereafter, we only need to consider contributions from the wall and the inside part.
The contribution of the wall is
\begin{eqnarray}
B_{\mathrm{wall}} = 2\pi^2 {\bar\rho}^3S_o,
\end{eqnarray}
where the surface tension of the wall, $S_o$, is a constant and $\bar\rho$ denotes the radius of the bubble \cite{Coleman:1980aw}.
The contribution from inside the wall is given by \cite{Parke:1982pm, Lee:2008hz}
\begin{eqnarray}
B_{\mathrm{in}} = \frac{12\pi^2}{\kappa^2} \left[
\frac{(1-\frac{\kappa}{3}V_{\mathrm{t}}\bar\rho^2)^{3/2}-1}{V_{\mathrm{t}}} -
\frac{(1-\frac{\kappa}{3}V_{\mathrm{f}}\bar\rho^2)^{3/2}-1}{V_{\mathrm{f}}} \right] \,.
\label{bin0}
\end{eqnarray}
If the strong gravity effect is considered\cite{Lee:2008hz}, one could obtain the relation
\begin{eqnarray}
V_{\mathrm{f}}-V_{\mathrm{t}} = \frac{3\kappa S_o^2}{4}
\end{eqnarray}
in the de Sitter background. The radius of the wall $\bar\rho$ can be obtained by extremizing $B$.
The above relation corresponds to
$12=\kappa \bar{\rho}^2_o (V_{\mathrm{f}}-V_{\mathrm{t}})$, where $\bar{\rho}_o(=3S_o/(V_{\mathrm{f}}-V_{\mathrm{t}}))$ denotes
the bubble radius in the absence of gravity.
If we take $V_{\mathrm{t}}=0$, the radius and the nucleation rate of a vacuum bubble can be simplified as
\begin{eqnarray}
\bar{\rho} = \frac{\bar{\rho}_o}{2},
\;\;\; B= \frac{B_o}{4} \,,
\label{case1-1}
\end{eqnarray}
where
\begin{eqnarray}
B_o = \frac{27\pi^2S_o^4}{2V^3_{\mathrm{f}}}
\end{eqnarray}
is the nucleation rate in the absence of gravity.

For the $O(3)$-symmetric bubble with gravitation, the nucleation rate and  radius of the bubble are
not clear when using the thin-wall approximation. It should be noted that a study has been conducted on the bubble
with $O(3) \times O(2)$ symmetry~\cite{Masoumi:2012yy}. Hereinafter, we adopt the results obtained from this study.
This one provides a distinguishable difference in the exponential suppression factor $B$ between two models,
one corresponds to the event without gravitation, and the other corresponds to the event with gravitation.
The inverse time duration of the phase transition corresponding to the nucleation rate of an $O(3)$-symmetric bubble
is given by \cite{Turner:1992tz, Caprini:2015zlo}
\begin{eqnarray}
\beta \equiv-\left.\frac{d B}{d t}\right|_{t=t_{*}} \simeq \frac{\dot{\Gamma}}{\Gamma} \,.
\end{eqnarray}
Hence, we obtain the following relation of $\beta$ at the time of the bubble collision:
\begin{eqnarray}
\beta_S \approx 4\beta_{S+G} \,,
\label{betasg}
\end{eqnarray}
where the subscript $``S"$ denotes the event without gravitation, while $``S+G"$ denotes
the event with gravitation. This result is one of the reasons why the GWs spectrum
can be distinguished between the scalar field only and that with gravity cases (Section~\ref{sec3}),
although the real value of the ratio will be smaller than $4$ times.

\subsection{Bubble collision \label{sec2-2}}

Now, we consider two identical vacuum bubbles with a preferred axis;
then, the geometry is reduced to the $O(2, 1)$ symmetry by the existence of the axis~\cite{Hawking:1982ga, Wu:1984eda, Kim:2014ara,Bond:2015zfa}.
To describe the bubble collision phenomenon, we adopt the double-null
formalism \cite{Hwang:2012pj, Nakonieczna:2018tih}. We choose the metric ansatz for the hyperbolic symmetry:
\begin{eqnarray}
ds^2= -\alpha^2(u,v)dudv+r^2(u,v)dH^2 \,,
\end{eqnarray}
where $r$ denotes a timelike coordinate, $dH^2= d\chi^2+\sinh^2\chi d\varphi^2$, while $u$ and $v$
correspond to the left- and right-going null directions, respectively. Note that typical black hole type solutions
correspond to $m < 0$; hence, we are interested in the case with $m < 0$. Accordingly,
we follow the procedure with some conventions and equations of motion used in \cite{Hwang:2012pj}.
The mathematical and numerical details of the simulations are summarized in Appendix.

In this paper, we briefly describe initial conditions along initial $u = 0$ and $v = 0$ surfaces.
\begin{description}
	\item[Initial $v=0$ surface:]
	we adopt
	\begin{eqnarray}
	S(u,0) = \left\{ \begin{array}{ll}
	{ {\bar \phi}_{\mathrm{f}}} & u < u_{\mathrm{shell}} \,,\\
	\left| { {\bar \phi}_t}-{ {\bar \phi}_{\mathrm{f}}} \right| G(u) + { {\bar \phi}_t} & u_{\mathrm{shell}} \leq u < u_{\mathrm{shell}}+\Delta u \,,\\
	{ {\bar \phi}_t} & u_{\mathrm{shell}}+\Delta u \leq u \,,
	\end{array} \right.
	\end{eqnarray}
	where $G(u)$ is a function that pastes from $1$ to $0$ in a smooth manner:
	\begin{eqnarray}
	G(u) = 1 - \sin^{2} \left[\frac{\pi(u-u_{\mathrm{shell}})}{2\Delta u}\right] \,.
	\end{eqnarray}
	\item[Initial $u=0$ surface:]
	We choose
	\begin{eqnarray}
	S(0,v) = \left\{ \begin{array}{ll}
	{{\bar \phi}_{\mathrm{f}}} & v < v_{\mathrm{shell}} \,,\\
	\left| { {\bar \phi}_{\mathrm{f}}} - { {\bar \phi}_t} \right| G(v) - { {\bar \phi}_t} & v_{\mathrm{shell}} \leq v < v_{\mathrm{shell}}+\Delta v \,,\\
	{ {\bar \phi}_t} & v_{\mathrm{shell}}+\Delta v \leq v \,,
	\end{array} \right.
	\end{eqnarray}
	where
	\begin{eqnarray}
	G(v) = 1 - \sin^{2} \left[\frac{\pi(v-v_{\mathrm{shell}})}{2\Delta v}\right] \,.
	\end{eqnarray}
\end{description}
%We use the second order Runge-Kutta method.

In order to specify a false vacuum background, we impose $r(0,0)=r_0$ and $S(0,0)={ {\bar \phi}_{\mathrm{f}}}$.
The two parameters $r_0$ and ${ {\bar \phi}_{\mathrm{f}}}$ are free parameters and the value of ${ {\bar \phi}_{\mathrm{f}}}$
can be considered as the field distance from the true to false vacuum. As a model parameter,
we consider ${ {\bar \phi}_{\mathrm{f}}=0.1}$ and ${0.2}$ in this work. In addition, we select $r_0= 1$ in the next calculation.
Regarding the detailed meanings of the boundary conditions and model parameters, refer to Appendix and \cite{Hwang:2012pj}.

\begin{figure}
\begin{center}
{\includegraphics[width=6.in]{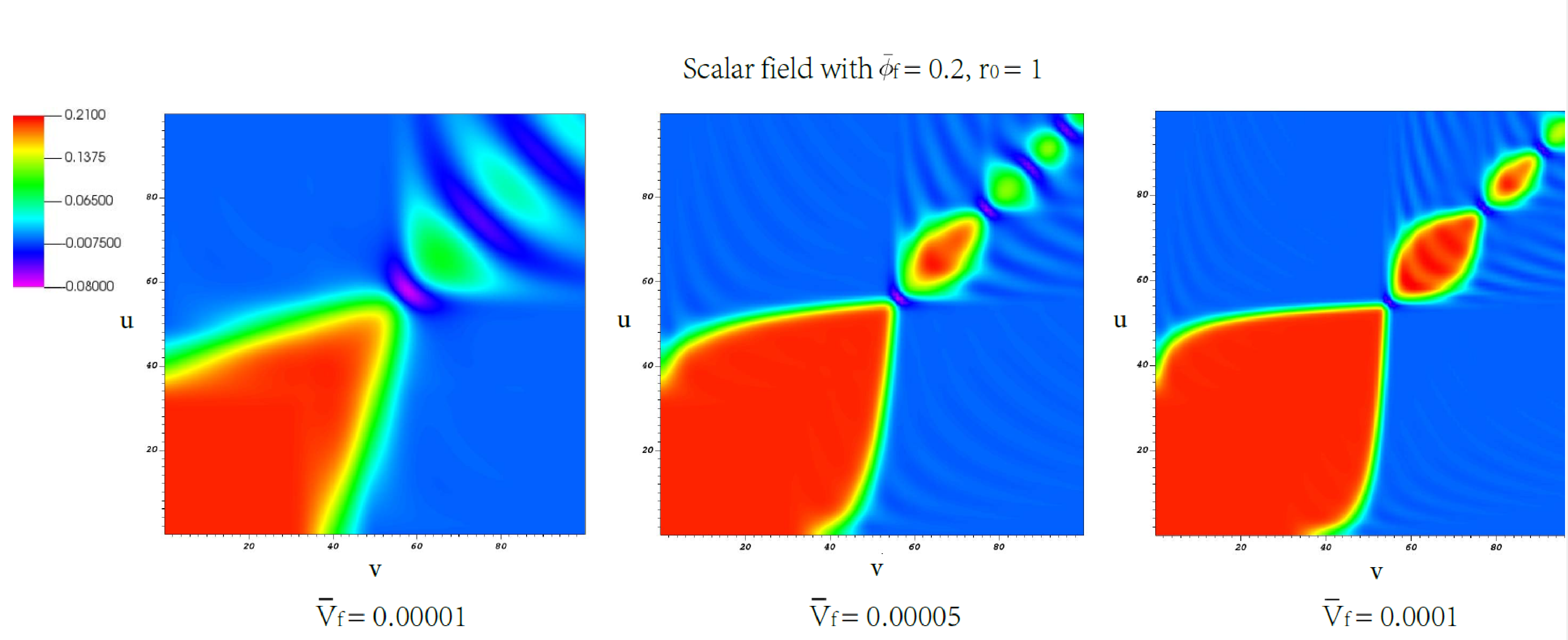}}
\end{center}
\caption{\footnotesize{(color online).
Bubble percolation with spherical symmetry: for ${{\bar \phi}_{\mathrm{f}}=0.2}$ and $r_0= 1$ at scalar field with
${ {\bar V}_{\mathrm{f}}=0.0.00001, 0.00005}$,
and ${0.0001}$, respectively.}}
\label{fg:1}
\end{figure}

\begin{figure}
\begin{center}
{\includegraphics[width=6.in]{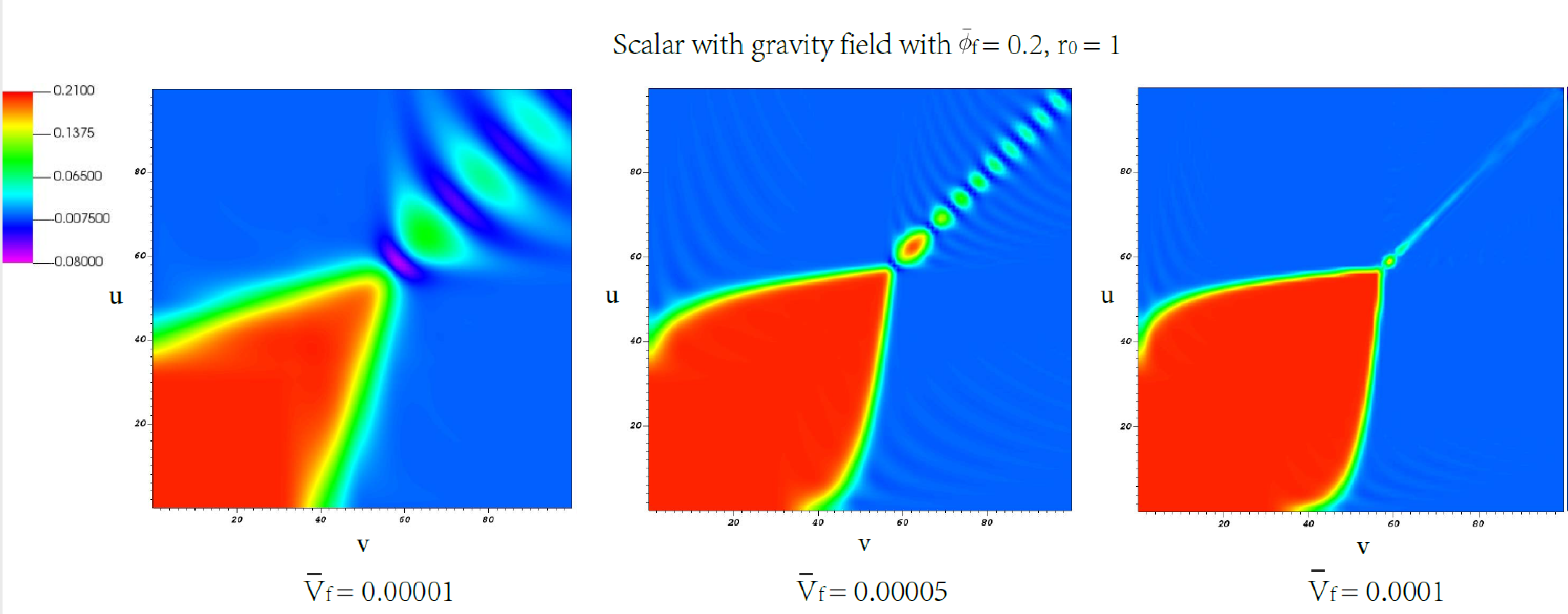}}
\end{center}
\caption{\footnotesize{(color online).
Bubble percolation with spherical symmetry:for ${ {\bar \phi}_{\mathrm{f}}=0.2}$ and $r_0= 1$ at scalar with gravity field with
${ {\bar V}_{\mathrm{f}}=0.00001, 0.00005}$,
and ${0.0001}$, respectively.}}
\label{fg:2}
\end{figure}

\begin{table}[ht]
	\begin{center}
		\caption{Velocity of bubble wall before collision in scalar with gravity field and scalar field with ${ {\bar \phi}_{\mathrm{f}}=0.2}$.}
		\begin{tabular}{|c|c|c|c|} \hline
			Value of $v_w$ & ${ {\bar V}_{\mathrm{f}}=0.00001}$ &${ {\bar V}_{\mathrm{f}}=0.00005}$&${ {\bar V}_{\mathrm{f}}=0.0001}$
			\\ \hline
			Scalar field& $0.895 $&$0.867$&$0.620$
			\\ \hline
			Scalar  with gravity field& $0.875$& $0.761$&$0.578$
			\\ \hline
		\end{tabular}
		\label{tab:2}
	\end{center}
\end{table}

Figures~\ref{fg:1} and \ref{fg:2} present the numerical results of the bubble collisions.
The red-colored region indicates large vacuum energy while the blue-colored region represents the region of low potential energy.
We can consider them as false vacuum and true vacuum regions, respectively.
As the wall oscillates, the oscillating field amplitudes decrease (upper right),
and eventually, the wall disappears (lower left). As the tension of the shell increases,
the initial expansion of the shell gradually slows down.
For the scalar field with the gravity case, the field oscillation becomes faster
than that of the scalar field without the  gravity case.

We also obtain the velocity of the bubble wall before collision from Figures \ref{fg:1} and \ref{fg:2}.
We approximately considered the slope of the yellow line and calculated the velocity by
\begin{eqnarray}
v_w =  \frac{\Delta \mathrm{v}- \Delta \mathrm{u}}{\Delta \mathrm{v}+ \Delta \mathrm{u}}.
\end{eqnarray}
The numerical results for $v_w$ in different potentials are presented in Table~\ref{tab:2}.

\begin{figure}
\begin{center}
{\includegraphics[width=3.5 in]{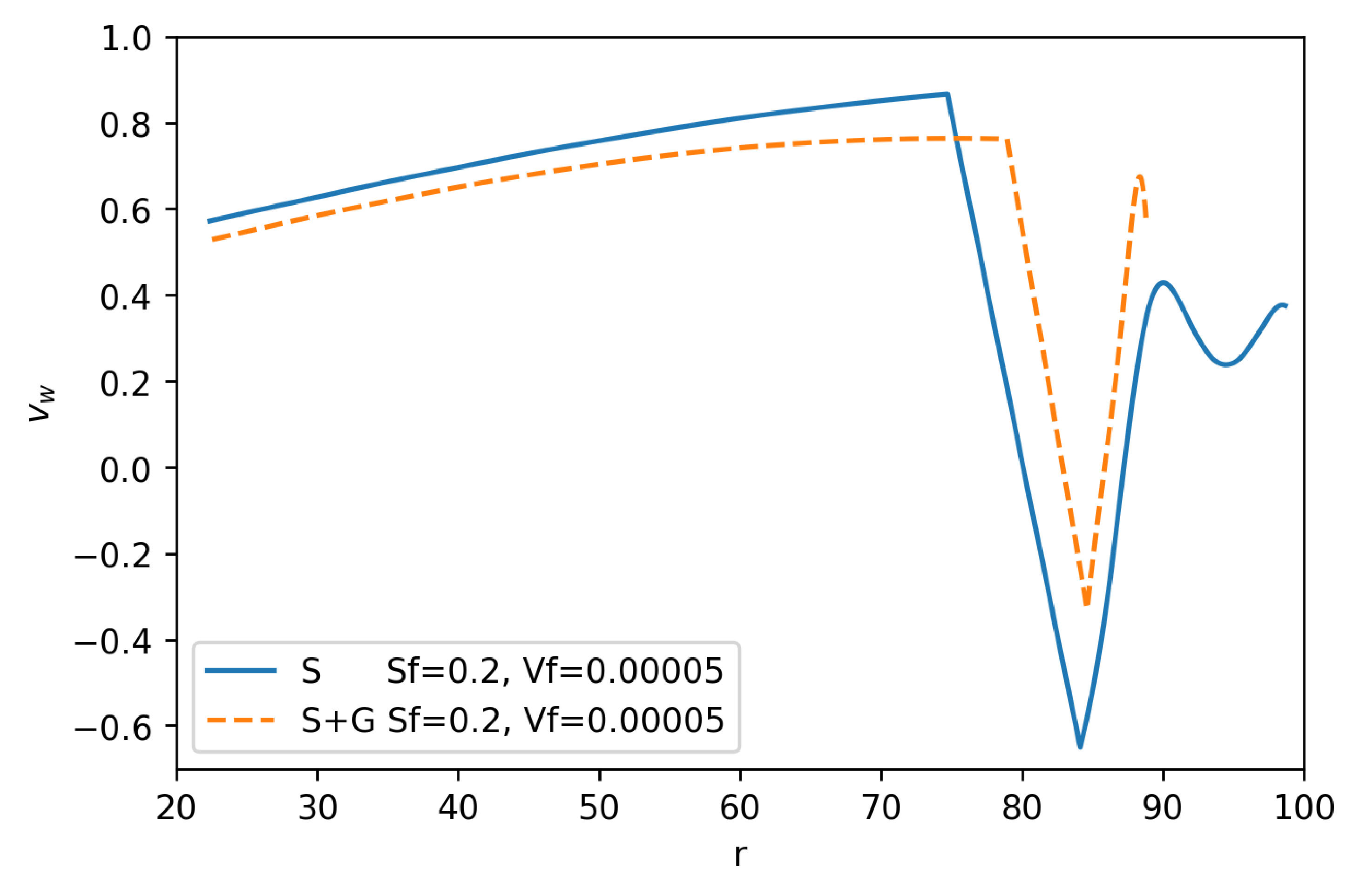}}
\end{center}
\caption{\footnotesize{(color online).
Time evolution of the bubble wall velocity. The $r$ is a timelike coordinate. The setting of potential are
same in scalar-only and scalar with gravity cases, which have
${ {\bar \phi}_{\mathrm{f}}=0.2}$, $r_0= 1$, and ${{\bar V}_{\mathrm{f}}=0.00005}$, respectively.}}
\label{fg:v}
\end{figure}

\begin{figure}
\begin{center}
\subfigure[]
{\includegraphics[width =2.5 in]{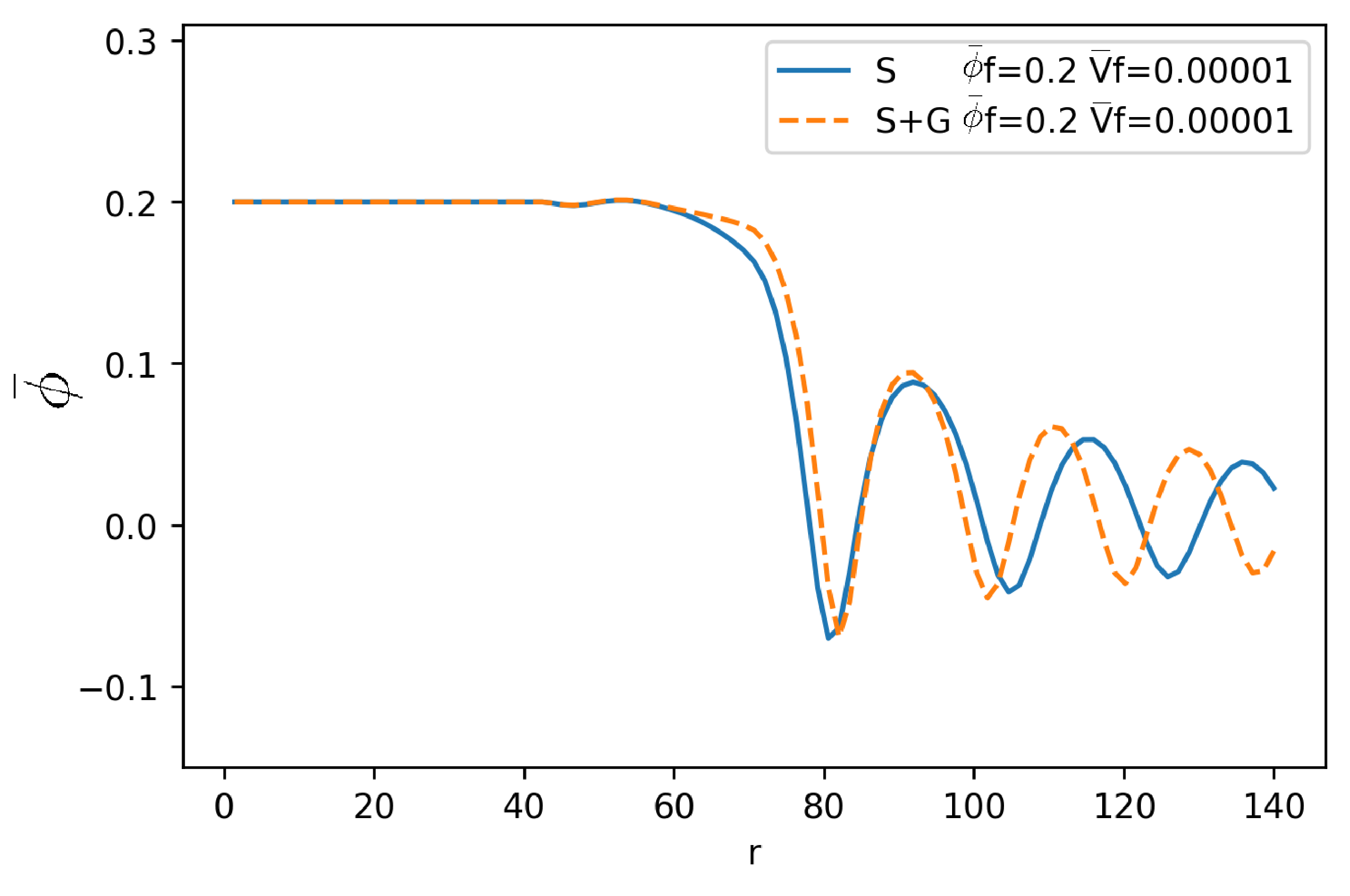}} 	\label{fg:a}
\subfigure[]
{\includegraphics[width =2.5 in]{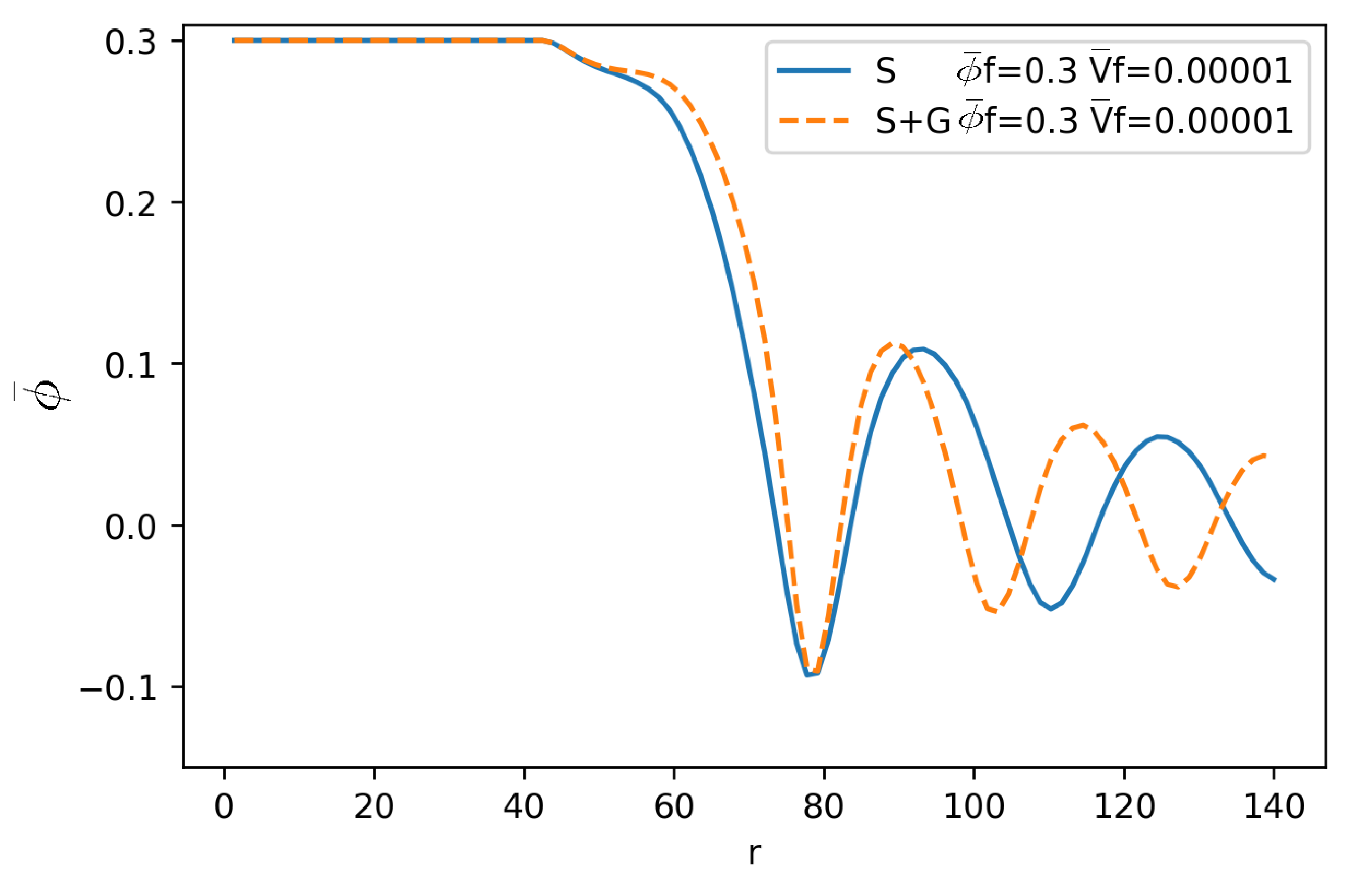}}   \label{fg:b}\\
\subfigure[]
{\includegraphics[width =2.5 in]{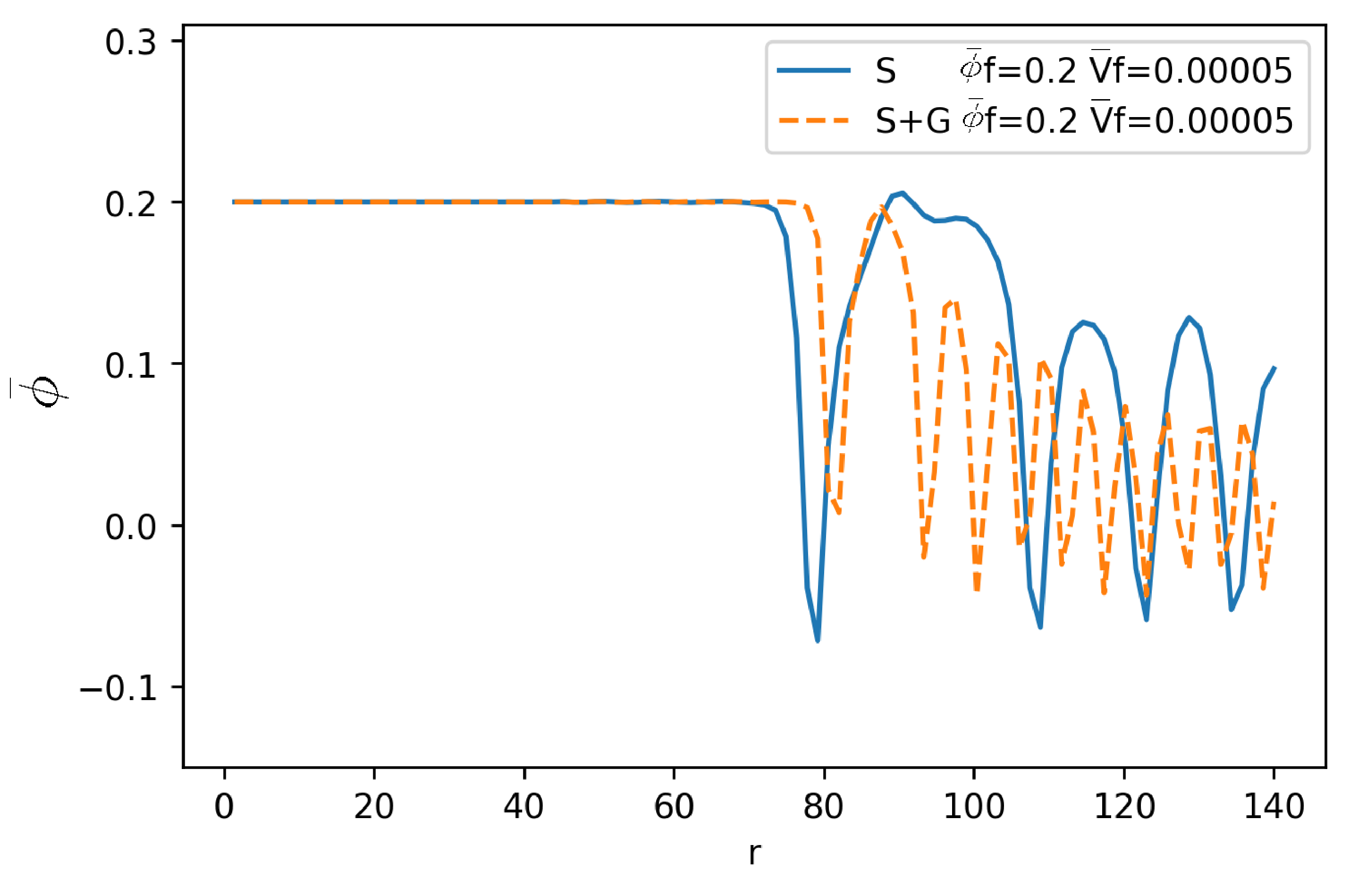}}   	\label{fg:c}
\subfigure[]
{\includegraphics[width =2.5 in]{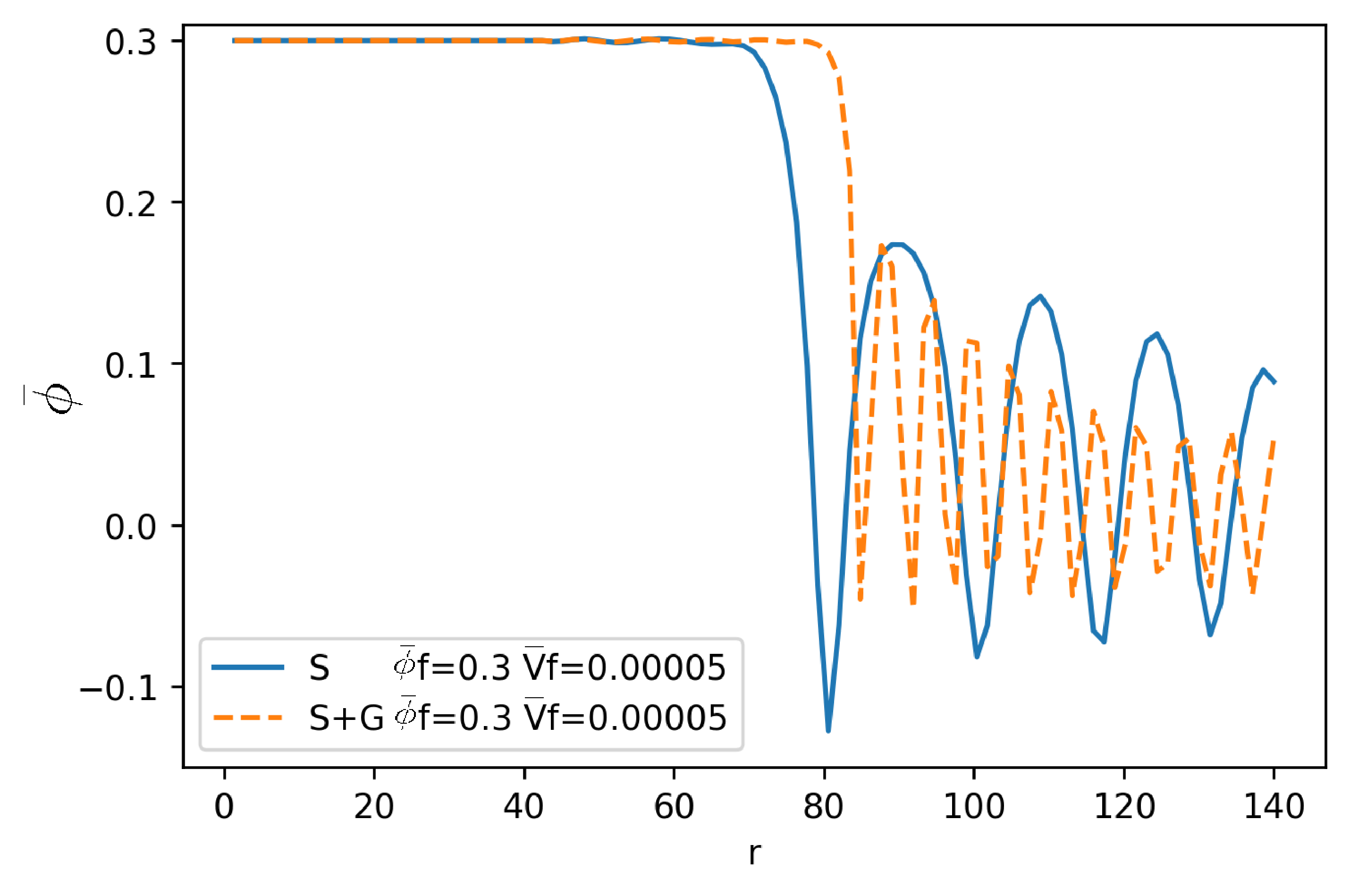}}   	\label{fg:d}
\end{center}
\caption{\footnotesize{(color online). Time evolution of the potential for one point from the false vacuum. The ``$S+G$" means the scalar with gravity field and the ``$S$" is the scalar field only case.}}
\label{fg:phi-t}
\end{figure}

The Figure~\ref{fg:v} presents the velocity evolution of the bubble wall for the scalar field
and that with the gravity cases. From the same set of potentials
(including ${ {\bar \phi}_{\mathrm{f}}=0.2}$ and ${ {\bar V}_{\mathrm{f}}=0.00005}$),
the scalar-only case has the faster velocity of the bubble wall than that with gravity case
before the bubble collision. The abscissa $r$ here is defined by $r= \mathrm{v} +\mathrm{u}$, which could not
correspond with real-time in cosmology. Before the collision, the numerical result of velocity
exhibits $v_w=0.867$ at approximately $r=75$ for the scalar field case, while $v_w$ equals to $0.761$ for
the case with gravity, as illustrated in Table~\ref{tab:2}. The label ``$S$" in the figure implies the scalar-only field case,
while the adopted ``$S+G$" represents the scalar with gravity case. Another information we can confirm from
this figure is that the case of ``$S+G$" exhibits a later collision time than the scalar-only case, and the frequency of the oscillation
is faster than that in the scalar-only case.

To understand the time evolution for the { vacuum state of the outside of the wall as the scalar field on the potential}, 
we select one point in the false vacuum and compare the oscillation of the point in different cases (refer to~\ref{fg:phi-t}).
When the two bubbles collide, the potential for this point will trun to the true vacuum. Then, the potential 
will oscillate back to zero. In  the same sub-figure of Figure~\ref{fg:phi-t}, the initial potential is also identical for the ``$S$" and ``$S+G$" cases. 
The comparative is interesting that the moment of two bubbles collision in ``$S+G$" will always be later than the scalar-only case.
In addition, the potential in the ``$S+G$" case will oscillate quicker than that in the scalar-only case.
We can compare the potential by different values of  ${ {\bar \phi}_{\mathrm{f}}}$ in (a) and (b) (also (c) and (d)).
For (a) and (c) (also (b) and (d)), we can observe the that the frequency of oscillation will significantly depend on the parameter ${ {\bar V}_{\mathrm{f}}}$.
The higher value of ${ {\bar V}_{\mathrm{f}}}$ will lead to a faster oscillation.
	
\section{Computation of gravitational wave spectrum \label{sec3}}

One of the most important assumptions for the GW spectrum from bubble collision is that GW does not depend
on the form of the potential~\cite{Kosowsky:1992vn}. In this section, we provide a comprehensive general
derivation process for the GW spectrum.

\subsection{Formalism of gravitational wave spectrum \label{sec3-1}}

Firstly, we allow the background spacetime to be dynamical, which implies that we would
like to define GWs as perturbations over some curved and dynamical
background metric ${\bar g}_{\mu\nu}$. Therefore, one can write
$g_{\mu\nu} = {\bar g}_{\mu\nu} +h_{\mu\nu}$ with $|h_{\mu\nu}| \ll 1$ \cite{Maggiore:1900zz}.
The ``coarse-grained" form of the Einstein equations can be given by
\begin{eqnarray}
{\bar R}_{\mu\nu} -\frac{1}{2}{\bar g}_{\mu\nu} {\bar R} =
\frac{8\pi G}{c^4} \left({\bar T}_{\mu\nu}+t_{\mu\nu} \right) \,,
\label{em1}
\end{eqnarray}
where the equation determines the dynamics of ${\bar g}_{\mu\nu}$, which is
the long-wavelength part of the metric. ${\bar T}_{\mu\nu}$ is the matter energy-momentum tensor
in terms of the long-wavelength, and the tensor ${t}_{\mu\nu}$ here does not depend on the external matter,
but on the gravitational field itself. These two tensor terms are also quadratic in terms of ${h}_{\mu\nu}$.

With the gauge invariant energy density in terms of the amplitudes ${ h}^2_+ $ and $ { h}^2_{\times}$ setting as
$t^{00} = \frac{c^2}{16\pi G} \langle{\dot h}^2_+ + {\dot h}^2_{\times}\rangle$,
we compute the corresponding energy flux straightforward. Because the energy-momentum tensor
can be expressed by the GWs, the total energy flowing through the surface of $dA$ is written as
\begin{eqnarray}
\frac{dE}{dA} = \frac{c^3}{16\pi G}\int^{+\infty}_{-\infty}dt \langle{\dot h}^2_+ + {\dot h}^2_{\times}\rangle \,.
\label{teneA}
\end{eqnarray}
In the next calculation, we focus on the determination of the GW spectrum.
Based on Eq.~(\ref{teneA}), the fundamental equation for the GW spectrum can be considered with
the angle $d\Omega$ as
\begin{eqnarray}
\frac{dE}{d\Omega} &=& \frac{r^2 c^3}{16\pi G}\int^{+\infty}_{-\infty}dt \left({\dot h}^2_+ + {\dot h}^2_{\times}\right) \\ \nonumber
  &=& \frac{G}{2\pi^2 c^7} \Lambda_{ij,kl}\left(\hat{\mathbf{k}}\right)\int^{+\infty}_0 d\omega \omega^2 {\tilde T}_{ij}\left(\omega, \frac{\omega\hat{\mathbf{k}}}{c}\right){\tilde T}^*_{kl}\left(\omega, \frac{\omega{\hat{\mathbf{k}}}}{c}\right) \,,
\label{teneOm}
\end{eqnarray}
where $d\omega$ indicates the frequency interval.
%After getting the stress-energy tensor with the frequency interval $d\omega$, t
The energy spectrum per solid angle is given by~\cite{Huber:2008hg}
\begin{eqnarray}
{\frac{d^{2}E}{d\omega d\Omega}}  =
2{G}{\omega^2} \Lambda_{ij,kl}\left({\hat{\mathbf{k}}}\right) {\tilde T}_{ij} \left({\hat{\mathbf{k}}},\omega\right){\tilde T}^*_{kl}\left({\hat{\mathbf{k}}}, \omega\right) \,,
\label{Om}
\end{eqnarray}
where $\Lambda_{ij,kl}$ in Eq.~(\ref{teneOm}) is the GWs' projection tensor,  and the full form is expressed as~\cite{Weinberg:1972kfs}:
\begin{eqnarray}
\Lambda_{ij,kl}\left({\hat{\mathbf{k}}}\right)= \delta_{il}\delta_{jm}- 2 {\hat{\mathbf{k}}}_j {\hat{\mathbf{k}}}_m \delta_{il} +\frac{1}{2}{\hat{\mathbf{k}}}_i {\hat{\mathbf{k}}}_j {\hat{\mathbf{k}}}_l {\hat{\mathbf{k}}}_m- \frac{1}{2}\delta_{ij}\delta{lm}+\frac{1}{2}{\hat{\mathbf{k}}}_l {\hat k}_m + \frac{1}{2}\delta_{lm}{\hat{\mathbf{k}}}_i {\hat{\mathbf{k}}}_j \,.
\label{lambda}
\end{eqnarray}

The instantaneously radiated power is significantly more useful because the total radiated energy is formally divergent.
In the Fourier space, the stress-energy tensor is
\begin{eqnarray}
T_{i j}\left(\hat{\mathbf{k}}, \omega\right)=\frac{1}{2 \pi} \int \mathrm{d} t
\mathrm{e}^{\mathrm{i} \omega t} \sum_{n} \int_{S_{n}} \mathrm{~d} \Omega \int \mathrm{d} r r^{2}
\mathrm{e}^{-\mathrm{i} \omega \hat{\mathrm{k}} \cdot\left(\mathrm{x}_{n}+r \hat{\mathrm{x}}\right)} T_{i j, n} \left(r, t\right)\,,
\label{Tij}
\end{eqnarray}
where the $\mathrm{x}_{n}$ is the position of the bubble in the nucleation, and $S_n$ is the un-collided region of the $n$-th bubble.

In the first-order phase transition, $1/\beta$ denotes the duration roughly, which will be set as the
frequency of GWs. Accordingly, we can write $\omega\approx\beta$ for the characteristic
frequency of the radiation. Meanwhile, we also consider the single bubble radius as $R$ in the below function,
and this can determine the scaling of the radiation spectrum amplitude:
\begin{eqnarray}
 \int^{R_n}_{0} \mathrm{d} r r^{2}  T_{i j}\left(r, t\right) \approx \frac{1}{3}  \hat{\mathrm{x}}_{i} \hat{\mathrm{x}}_{j} R_{n}^{3} \kappa_{\phi} \epsilon
 = \frac{1}{4}  \hat{\mathrm{x}}_{i} \hat{\mathrm{x}}_{j} R_{n}^{3} \kappa_{\phi} \alpha \omega_1,
\end{eqnarray}
where $\kappa_{\phi}$ is the efficiency factor introduced previously
and the subscript $1$ denotes the quantity in the symmetric phase~\cite{Kamionkowski:1993fg}.

We ignore the $\mathrm{e}^{-\mathrm{i} \omega \hat{\mathrm{k}} \cdot\left(\mathrm{x}_{n}+r \hat{\mathrm{x}}\right)}$ term as $1$ in Eq.~(\ref{Tij}) and consider the length scale as $R\simeq v/\beta$. With the number of bubble setting as $N$, we can obtain the approximation of  GW spectrum in the bubble collision time as
\begin{eqnarray}
\frac{dE}{d\omega} \varpropto N G \left(R^3\kappa_{\phi}  \epsilon\right)^2.
\end{eqnarray}
In addition, this equation can be simplified as ${\frac{1}{ E_{vac}} \frac{dE}{d\omega}} \varpropto  G v^3 \kappa_{\phi}^2 \alpha \frac{\omega_1}{\beta^3}.$

The fraction of energy liberated into the GW radiation per frequency octave is
\begin{eqnarray}
\Omega_{\mathrm{GW} *}=\omega \frac{\mathrm{d} E_{\mathrm{GW}}}{\mathrm{d} \omega} \frac{1}{E_{\mathrm{tot}}}=\kappa_{\phi}^{2}\left(\frac{H_*}{\beta}\right)^{2}\left(\frac{\alpha}{\alpha+1}\right)^{2} \Delta\left(\omega / \beta, v_{\mathrm{b}}\right),
\end{eqnarray}
where $E_{vac}$ denotes the total vacuum energy in the sample volume, which is given
as $E_{vac} \simeq N R^3 \epsilon$ and the $\Delta$ is a dimensionless function term, which can be defined as
$\Delta\left(\frac{\omega}{\beta}, v_{\mathrm{b}}\right)=\frac{\omega^{3}}{\beta^{3}} \frac{3 v_{\mathrm{b}}^{6}
\beta^{5}}{2 \pi V} \int \mathrm{d} \hat{\mathrm{k}} \Lambda_{i j, l m} C_{i j}^{*} C_{l m}$, in which
one can refer to Ref. \cite{Huber:2008hg} for the definition of $C_{ij}$.
The star subscript ($*$) refers to the means at the time of the first order phase transition happened.
The Hubble parameter at the time of the phase transition is
\begin{eqnarray}
H_{*}^{2}=\frac{8 \pi G \rho_{\mathrm{rad}}}{3}=\frac{8 \pi^{3} g_{*} T_{*}^{4}}{90 m_{\mathrm{Pl}}^{2}}.
\end{eqnarray}
One can obtain $\int \mathrm{d} \hat{\mathbf{k}} \Lambda_{i j, l m} C_{i j}^{*} C_{l m}
\propto {N \beta^{-8} \propto \frac{V}{v_{\mathrm{b}}^{3} \beta^{5}}}$ can be obtained in the quadrupole approximation,
if the integration is performed over a large volume $V$ of the Universe.

After obtaining the spectrum of  $\Omega_{\mathrm{GW} *}$ in the moment of bubble collision,
we are going to calculate the corresponding signal in the present Universe.
We consider that the entropy per comoving volume is always the same in the history of the Universe, and
the relationship can be set as $S \varpropto R^3 g(T) T^3$. At the temperature $T_*$,
the number of relativistic degrees of freedom was expressed in $g(T_*)$. Hence, we can obtain
\begin{eqnarray}
\frac{R_{*}}{R_{0}}=8.0 \times 10^{-14}\left(\frac{100}{g_{*}}\right)^{1 / 3}\left(\frac{1 \mathrm{GeV}}{T_{*}}\right),
\end{eqnarray}
where $g_*= 106.75$ was considered at an electroweak scale.

The $f_*$ parameter denotes the characteristic frequency at the transition, and the characteristic frequency
$f_0$ today can be considered as \cite{Huber:2008hg}
\begin{eqnarray}
f_{0}=f_{*}\left(\frac{R_{*}}{R_{0}}\right)=1.65 \times 10^{-7} \mathrm{~Hz}\left(\frac{f_{*}}{H_{*}}\right)\left(\frac{T_{*}}{1 \mathrm{GeV}}\right)\left(\frac{g_{*}}{100}\right)^{1 / 6},
\end{eqnarray}

\begin{eqnarray}
\Omega_{\mathrm{GW}} h^{2}=\Omega_{\mathrm{GW} *}\left(\frac{R_{*}}{R_{0}}\right)^{4}H_{*}^{2}
=1.67 \times 10^{-5} \left(\frac{100}{g_{*}}\right)^{1 / 3} \Omega_{\mathrm{GW} *},
\end{eqnarray}
where $h$ is the current Hubble parameter result.
Note that the quantity $\Omega_{\mathrm{GW}}h^2$ is independent of the actual Hubble expansion rate.
Hence, that quantity is often adopted, rather than $\Omega_{\mathrm{GW}}$ alone.

We can parameterize the spectrum of the GWs (the energy density spectrum)
from bubble collisions by
\begin{eqnarray}
\Omega_{\mathrm{GW}}(f) h^{2}
=1.67 \times 10^{-5} \left(\frac{100}{g_{*}}\right)^{1 / 3} \left(\frac{H_*}{\beta}\right)^{2}
\left(\frac{\kappa_{\phi} \alpha}{\alpha+1}\right)^{2} \Delta S(f)
\end{eqnarray}
where the parameter $\beta/H_*$ controls the GW signal. A small value of $\beta/H_*$ means the fast
first-order phase transition.
$\Delta =\frac{0.11 v_b^3}{0.42+v_b^2}$ is obtained from the numerical result in \cite{Huber:2008hg} and
$S(f)= \frac{3.8(f/f_0)^{2.8}}{1+2.8(f/f_0)^{3.8}}$
parameterizes the spectral shape of the GW radiation.

\subsection{Gravitational wave spectrum with LISA \label{sec3-2}}

{In this study, we solely focus on the contribution of the scalar field without the sound wave and turbulence.
%Thus there is no sound wave and turbulence analyzed in \cite{Caprini:2015zlo,Schmitz:2020rag}.
We compare the GW spectra both from the scalar field model and that with
gravity by the LISA sensitivity.
%If the spectrum from the scalar field only model can be contained in the sensitivity of eLISA,
%the real GW spectrum are no doubt to be observed within the sound wave and turbulence.
The bubble collisions will produce a stochastic background of GWs.
The stochastic backgrounds are random GW signals due to the superposition of
many independent uncorrelated sources that cannot be resolved individually~\cite{Hogan:1986qda},
which can be tested by the space-based interferometer LISA.
The sensitivity result of LISA for a stochastic gravitational-wave background was also presented
in Figure~\ref{fg:GW}. The redline represents the sensitivity curve derived from the noise model we adopt in LISA.We consider the sensitivity of LISA with several parapmeters{\cite{Caprini:2019pxz,Robson:2018ifk}}. We used an observation time of 4 years with 75$\%$ efficiency, meaning that we set T = 3 years.
In the figure, the Signal to Noise Ratio (SNR) was  in fact at approximately
7, which confirms that the choice SNR $\approx$ 10 for the gravitational wave signal is feasible, as depicted by the purple line and also found in {\cite{Adams:2013qma}}.
%which is based on $6$ links, $5$ million km arm length, and a duration of five years.
%Its noise level is $N2$ which can be called ``LISA Psathfinder expected".
%The $C4$ line shows the least sensitive result and it is the lowest detection threshold,
%which is based on $4$ links, $1$ million km arm length, and a duration of two years' observation.
%The noise level of $C4$ is $10$ times larger than that in $C1$, so it is named ``LISA Pathfinder required".
%The $C4$ can be the first result obtained from eLISA. In the next discussion,
%we can compare $C1$ (as eLISA result for five years) and $C4$ (as eLISA result for two years)
%with the theoretical sensitivity of the bubble collision directly.

\begin{figure}
\begin{center}
\subfigure[]
{\includegraphics[width =2.5 in]{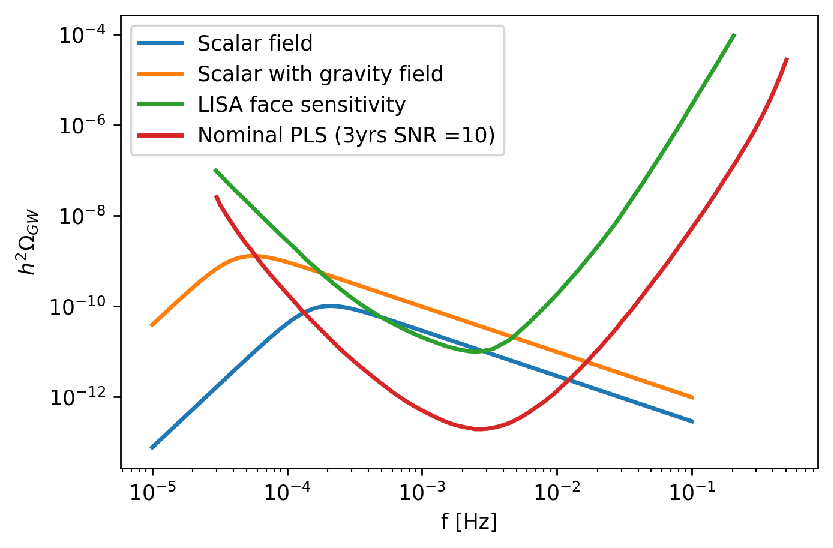}} 	
\subfigure[]
{\includegraphics[width =2.5 in]{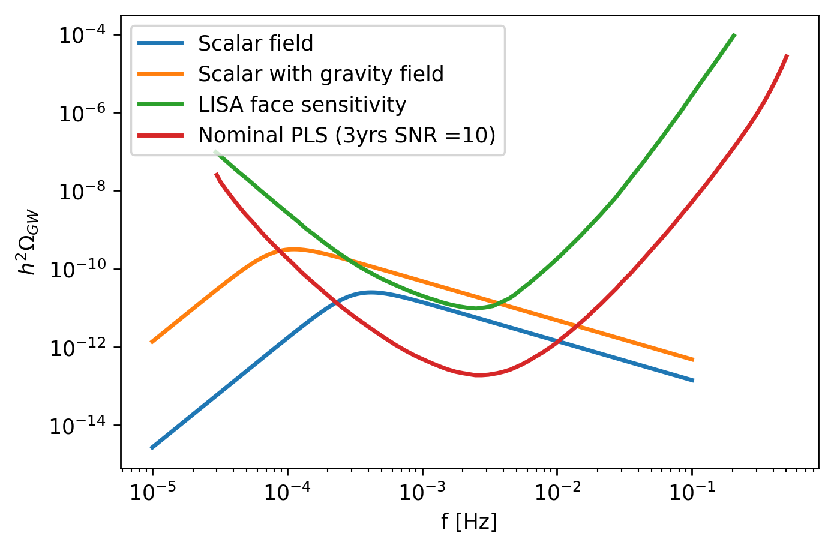}}   \\
\subfigure[]
{\includegraphics[width =2.5 in]{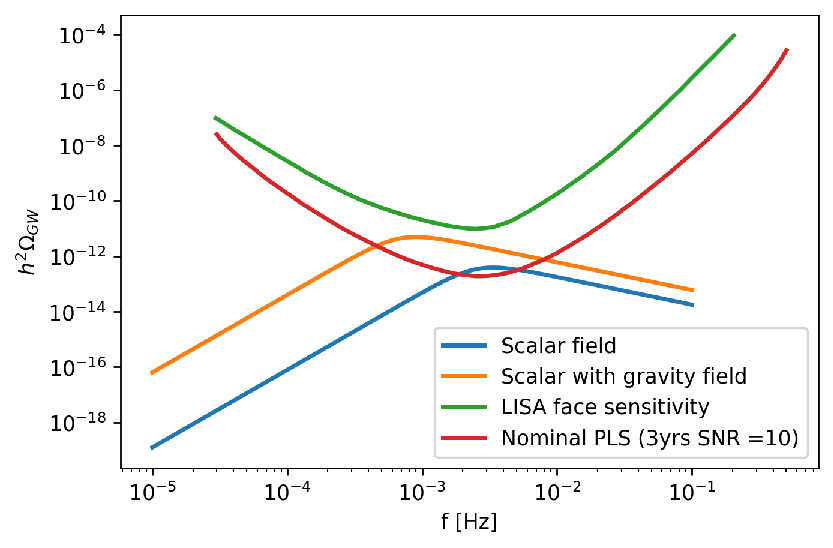}}
\subfigure[]
{\includegraphics[width =2.5 in]{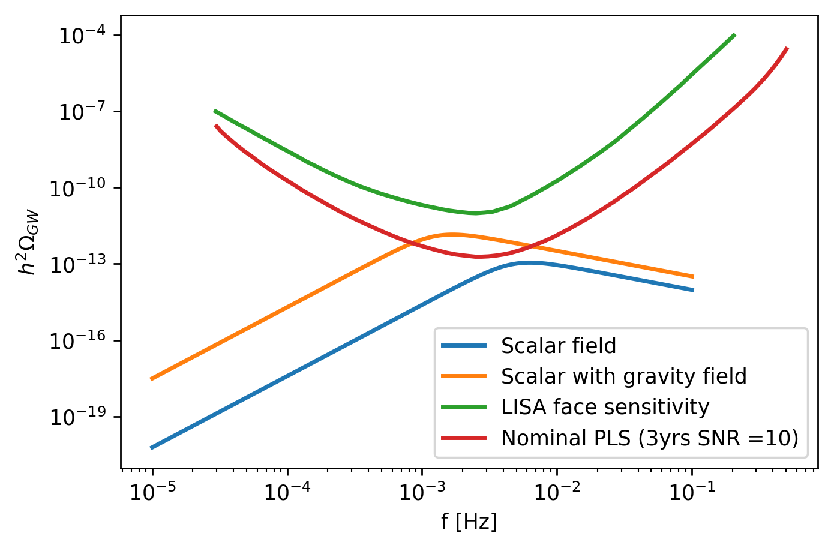}}
\end{center}
\caption{\footnotesize{(color online). Gravitational wave spectrum considered
the contribution from bubbles with ${ {\bar \phi}_{\mathrm{f}}=0.2}$ and ${ {\bar V}_{\mathrm{f}}=0.00005}$.
The value of $\beta_{s}/H_*$ are $50$, $100$, $800$, and $1500$, respectively.
The broken power law sensitivity curve derived from the noise model we adopt (green curve) and the corresponding
PLS for T = 3 yr with SNR= 10 (red curve)
 $v_w^S=0.867$ and $v_w^{S+G}=0.761$ in this figure.}}
\label{fg:GW}
\end{figure}

The comparison of results can be PRESENTED in Fig.\ref{fg:GW}. In this figure, the  $\beta_{s}/H_*$ values
are $50$, $100$, $800$, and $1500$, respectively. Meanwhile, the corresponding values for cases with gravity
are $12.5$, $25$, $200$, and $375$, which are the same as the relationship in Eq. (\ref{betasg}).
The larger value of $\beta_{s}/H_*$ implies the relatively slow first-order phase transition.
In slower first-order phase transitions, most of the vacuum energy is released into
the acceleration of the bubble walls \cite{Cai:2017tmh}.
The value of $T_*$ is 100 GeV, and for $v_w$, we adopted the case for ${ {\bar \phi}_{\mathrm{f}}=0.2}$ and ${ {\bar V}_{\mathrm{f}}=0.00005}$.
%The red line is the eLISA's five years of sensitive results, and the purple line means
%the sensitive result of eLISA's two years of observation.

These results are easy to read from Figure~\ref{fg:GW}.
First, the GW spectrum for the scalar with gravity case has a larger amplitude than that in the scalar-only
case at the same frequency.
{ When gravity is considered, the peak frequency shifts to the upper left.}
 In particular, when $\beta_{s}/H_*=1500$, the $S+G$ case can be tested
by the LISA observation; however, the scalar-only case exhibits our  range. This suggests that the scalar with the gravity case will be  tested easier than
the scalar-only case in the same original setting, which implies
the possibility for distinguishing the two cases in the LISA observation.
%Secondly, the spectrum from the scalar case has been out of range of the eLISA sensitivity
%when $\beta_{s}/H_*=2000$, but the amplitude of the scalar with gravity case still can be observed.
This result is mainly based on the different contributions from the $\beta$ parameter.
}
\section{Summary and discussions \label{sec4}}

In this study, we investigated the gravitational wave spectrum originating from the bubble collision and
compared the difference of the parameters ${{\bar \phi}}$, $\beta$, and $v_w$ in the scalar-only field
and that with gravity field cases.
We compared the evolution of $v_w$ from numerical simulation.
In the same initial distance, the $v_w$ in scalar with gravity case will always be smaller than the scalar-only case, and $v_w$ will also be reduced with a larger potential energy difference.
 The ratio $\beta_S/\beta_{S+G}$ is approximately $4$ times
in the Minkowski space, and this difference in $\beta$ triggered the main differences in the peak and
the amplitude of { the} gravitational wave spectrum. In the comparison of gravitational wave spectrum produced
from the two fields in the vacuum-only state, we can observe that the amplitude in the scalar field case with gravity
can exhibit have a larger value than that in the scalar-only field case.  With this higher amplitude, we expected {
that the LISA with SNR=10 could observe the spectrum as the fast first-order phase transition.}
%or that with $C1$ as the slow first-order phase transition from the latter case.
%even have the possibility to be observed in only two years' eLISA observation.
Compared to previous studies that did not consider the effect of gravitation,
our present study considered the effect of gravitation. At the electroweak phase transition,
it may be believed that the effect was not significant, but we considered the case when the effect was maximum.
{  Hence, we expect that the gravitational wave spectrum will be relatively above, but not significantly in line without the gravitational effect. 
%However, that will be between the line without the gravitational effect and that with the strong one. 
However, that will be between the line without the gravitational effect and that with the strong one.
We would like to emphasize that we have provided an upper bound in this paper, 
even though the gravitational effect was negligible at the electroweak phase transition era.
}

In one of our previous works, we investigated gravitational waves from cosmic bubble collisions,
in which time-domain gravitational waveforms were directly obtained
by integrating the energy-momentum tensors over the volume
of the wave sources~\cite{Kim:2014ara}.
Both the time-  and
 frequency-domains of the gravitational wave analysis in the context of the bubble collisions
might be interesting not only theoretical but also from the experimental perspective. Based on this work, along with the previous work,
it is believed that they will be in a mutually complementary relationship to investigate the
cosmological gravitational wave spectrum.

It will also be interesting to study bubble collisions not only in Einstein's gravity but also
in various other models including string-inspired scenarios \cite{Hwang:2013zaa,Hansen:2014rua,Hansen:2015dxa}
or modified gravity models \cite{Hwang:2010aj,Hwang:2011kg,Chen:2015nma}.
We leave these possible directions for future research topics.

\section*{Acknowledgments}

BHL (2020R1F1A1075472), DY (2021R1C1C1008622,
 2021R1A4A5031460),
  and WL (2016R1\\D1A1B01010234) were supported by the National Research Foundation of Korea. LY was supported by the Basic Science Research Program (2020R1A6A1A03047877) of the National Research Foundation of Korea funded by the Ministry of Education through Center for Quantum Spacetime (CQUeST) of Sogang University.
  {   The authors thank the organizers for their hospitality at the 17th
  	Italian-Korean Symposium for Relativistic Astrophysics in Kunsan National University, 02-
  	06 August 2021.} The authors also thank Hyeong-Chan Kim and Youngone Lee for their hospitality during our visit to Korea National University of Transportation;
Hoon Soo Kang, Keun-Young Kim, Chanyong Park, Yun-Seok Seo, and Hyun Seok Yang to GIST; Seoktae Koh to Jeju National University.
BHL is grateful to APCTP for their hospitality during his visit.
We appreciate the discussion from Gianmassimo Tasinato and Ke-Pan Xie.
\appendix
\section*{Appendix: brief review of the double-null simulation}

The generic review of the double-null simulation is in Ref.~\cite{Nakonieczna:2018tih, Hong:2008mw}.
For  convenience, we first define
\begin{eqnarray}
\sqrt{4\pi}\phi \equiv S.
\end{eqnarray}
In addition, we define 
\begin{eqnarray}\label{eq:conventions}
h \equiv \frac{\alpha_{,u}}{\alpha},\quad d \equiv \frac{\alpha_{,v}}{\alpha},\quad f \equiv r_{,u},\quad g \equiv r_{,v},\quad W \equiv S_{,u},\quad Z \equiv S_{,v},
\end{eqnarray}
  to present all equations as a set of first-order partial differential equations. In particular, for examples to study vacuum bubbles in double-null simulations, refer to Refs.~\cite{Hansen:2009kn,Hwang:2010gc,Hwang:2012nn}.

After a simple derivation, we obtain the Einstein tensor components
\begin{eqnarray}
\label{eq:Guu}G_{uu} &=& -\frac{2}{r} \left(f_{,u}-2fh \right),\\
\label{eq:Guv}G_{uv} &=& \frac{1}{2r^{2}} \left( 4 rf_{,v} - \alpha^{2} + 4fg \right),\\
\label{eq:Gvv}G_{vv} &=& -\frac{2}{r} \left(g_{,v}-2gd \right),\\
\label{eq:Gthth}G_{\chi\chi} &=& -4\frac{r^{2}}{\alpha^{2}} \left(d_{,u}+\frac{f_{,v}}{r}\right),
\end{eqnarray}
and the energy-momentum tensor components
\begin{eqnarray}
\label{eq:TPhiuu}T^{\phi}_{uu} &=& \frac{1}{4 \pi} W^{2},\\
\label{eq:TPhiuv}T^{\phi}_{uv} &=& \frac{\alpha^{2}}{2} V(S),\\
\label{eq:TPhivv}T^{\phi}_{vv} &=& \frac{1}{4 \pi} Z^{2},\\
\label{eq:TPhithth}T^{\phi}_{aa} &=& \frac{r^{2}}{2 \pi \alpha^{2}} WZ -r^{2}V(S).
\end{eqnarray}
Therefore, the Einstein equations are as follows:
\begin{eqnarray}
\label{eq:E1}f_{,u} &=& 2 f h - 4 \pi G r T^{\phi}_{uu},\\
\label{eq:E2}g_{,v} &=& 2 g d - 4 \pi G r T^{\phi}_{vv},\\
\label{eq:E3}f_{,v}=g_{,u} &=& - \kappa \frac{\alpha^{2}}{4r} - \frac{fg}{r} + 4\pi G r T^{\phi}_{uv},\\
\label{eq:E4}h_{,v}=d_{,u} &=& -\frac{2\pi G \alpha^{2}}{r^{2}}T^{\phi}_{aa} - \frac{f_{,v}}{r}.
\end{eqnarray}
Here, $G$ is the gravitational constant; if we take $G = 0$, this corresponds to the scalar-only field case without gravitational back-reactions. The scalar field equation becomes
\begin{eqnarray}
\label{eq:S}Z_{,u} = W_{,v} = - \frac{fZ}{r} - \frac{gW}{r} -\pi\alpha^{2} V'(S).
\end{eqnarray}

By comparing  with the static metric, we can determine a principle to provide the physical boundary condition. The static metric form is
\begin{eqnarray}
ds^{2} = - N^{2}(r) dt^{2} + \frac{dr^{2}}{N^{2}(r)} + r^{2} dH^{2},
\end{eqnarray}
where
\begin{eqnarray}
N^{2} = - \left( 1 + \frac{2M}{r} + \frac{8\pi V r^{2}}{3} \right).
\end{eqnarray}
For typical hyperbolically symmetric cases, $t$ is the space-like parameter and $r$ denotes the time-like parameter. By comparing to the double-null metric, we obtain
\begin{eqnarray}
N^{2} = -\frac{4r_{,u}r_{,v}}{{\alpha}^{2}}.
\end{eqnarray}
Hence, one can choose $r_{,u}>0$ and $r_{,v}>0$. The Misner-Sharp mass function in the double-null coordinate is
\begin{eqnarray}
m(u,v) = -\frac{r}{2} \left( 1 - \frac{4 r_{,u} r_{,v}}{\alpha^{2}} + \frac{8\pi V}{3} r^{2} \right).
\end{eqnarray}
Note that usual black hole type solutions in Minkowski vacuum will occur for $m < 0$ limit; hence, in this study, we are interested in $m \leq 0$.

Finally, boundary conditions along initial $u=0$ and $v=0$ surfaces are assigned as follows.
\begin{description}
\item[$v=0$ surface:]
We first choose $S(u,0)$. Then, we know $W(u,0)=S_{,u}(u,0)$. From this, $h(u,0)$ is given from Equation~(\ref{eq:E1}), since $f_{,u}=0$ along the in-going null surface. Then, using $h(u,0)$, we obtain $\alpha(u,0)$. In addition to them, we further obtain $d$ from Equation~(\ref{eq:E4}), $g$ from Equation~(\ref{eq:E3}), and $Z$ from Equation~(\ref{eq:S}).
\item[$u=0$ surface:]
we first adopt $S(0,v)$. Then, we obtain $d(0,v)$ from Equation~(\ref{eq:E2}),as $g_{,v}(0,v)=0$. By integrating $d$ along $v$, we have $\alpha(0,v)$. In addition, we further obtain $h$ from Equation~(\ref{eq:E4}), $f$ from Equation~(\ref{eq:E3}), and $W$ from Equation~(\ref{eq:S}).
\end{description}
We used the $2$nd-order Runge-Kutta method. The consistency and the convergence checks were analyzed in Ref.~\cite{Hwang:2012pj}.

\end{document}